\documentclass[conference]{IEEEtran}
\ifCLASSINFOpdf
   \usepackage[pdftex]{graphicx}
\else
\fi
\usepackage{amsmath}
\usepackage{algorithmic}

\usepackage{array}

\usepackage{stfloats}

\usepackage{cite}

\usepackage{fancyhdr}
\begin{document}

%
\title{A Wideband Sliding Correlator-Based Channel Sounder with Synchronization in 65 nm CMOS}

\author{
Ting Wu, Theodore S. Rappaport, Michael E. Knox, Davood Shahrjerdi\\

\IEEEauthorblockA{NYU WIRELESS, NYU Tandon School of Engineering, New York University, Brooklyn, NY, 11201\\}

Email:\{tw1059, tsr, mek407, davood\}@nyu.edu}



%


\maketitle
\thispagestyle{fancy}

\begin{abstract}
 A programmable ultra-wideband sliding correlator-based channel sounder with high temporal and spatial resolution is designed in standard 65 nm CMOS.  The baseband chip can be configured either as a baseband transmitter to generate a pseudo-random spread spectrum signal with flexible sequence lengths, or as a baseband receiver with sliding correlator having an absolute timing reference to obtain power delay profiles of the multipath components of the wireless channel. The sequence achieved a chip rate of one Giga-bit-per-second, resulting in a multipath delay resolution of 1 ns. The baseband chip occupies an area of 0.66 mm x 1 mm with a power dissipation of 6 mA at 1.1 V in 65 nm CMOS. The sliding correlator-based channel sounder in this work is a critical block for future low-cost, miniaturized channel sounding systems used in accurate and efficient channel propagation measurements at millimeter-wave frequencies. 
\end{abstract}


%
\IEEEpeerreviewmaketitle

\section{Introduction}
Advances in manufacturing of silicon integrated circuits have enabled wideband wireless communications networks that operate at millimeter-wave frequencies~\cite{hashemi2016mm,rappaport2014millimeter}. These higher frequency bands have massive amounts of raw bandwidth, thereby allowing remarkably high data rates (tens of Giga-bit-per-second, (Gbps)) in spectrum bandwidths that are several hundreds of MHz to a few GHz. A critical step towards the implementation of these emerging wireless technologies is the development of accurate channel propagation models for the frequency bands of interest. 


The sliding correlation measurement technique has provided a gold standard for accurate modeling of channel propagation. This direct sequence spread spectrum technique, pioneered by Don Cox of Bell Telephone Labs~\cite{Cox1972:1}, performs channel propagation measurements in the time domain by transmitting an RF carrier modulated with a high chip rate pseudo-random code and correlating the received signal with a replica of this PN code at the receiver. The output from the correlator represents the power delay profile (PDP) of the wireless channel. Past research has successfully demonstrated the utility of this technique for characterizing radio channels with RF carrier operating at millimeter-wave frequencies up to 73 GHz~\cite{WillWork:TSR13,Zwick2005:1, rappaport2015wideband,samimi201628,maccartney2017flexible,ben2011millimeter,rappaport2013broadband,rappaport201238,anderson2002building,hur2014synchronous,lee2015directional,miao2015indoor,Mac17JSACb}. Despite the progress for making propagation measurements at such high frequencies, the increase of the RF bandwidth has remained a difficult task as the pseudo-random sequence requires very high chip rates in order to achieve high resolution in the PDP. Given the short wavelength of the signal at these high carrier frequencies, the signal would experience significant scattering \cite{ju19icc}. Therefore, to improve the accuracy of the propagation channel models, it is essential to increase the RF bandwidth. The RF bandwidth of the most current measurement systems is still below 1 GHz~\cite{maccartney2017flexible,lee2015directional,miao2015indoor}. The aim of our paper is to report a chip design that overcomes this limit using low-cost CMOS technology.   

The hardware implementation of most sliding correlators reported in the literature have relied on commercial off-the-shelf (COTS) electronic components \cite{anderson2002building,maccartney2017flexible,lee2015directional,miao2015indoor,Mac17JSACb}. However, the use of COTS for implementing a practical high resolution measurement systems has a few shortcomings. First, the limitation in the maximum chip rate of the baseband electronics reduces the resolution in the power delay profile. Second, COTS-based implementations are often complex and relatively bulky, hindering the ability to make measurements in tight spaces, including crowded offices, inside vehicles and carried by an individual. Third, as massive multiple-input-multiple-out (massive MIMO) is becoming a key research subject for next-generation cellular networks, there is also a need for multiple transmitting and receiving channels that work in parallel to provide a short measurement time. However, COTS hinders the implementation of multiple sounding and receiving channels \cite{sun2014mimo}. Lastly, it increases the overall cost of the implementation, which may be prohibitively expensive for many research entities. 

To mitigate the aforementioned problems, we present a low-cost, ultra-wideband, programmable channel sounder with synchronization capability, implemented in a 65 nm standard CMOS process. We anticipate that our wideband channel sounder baseband integrated circuit (IC) will improve the efficiency and accuracy of the channel propagation measurements at millimeter-wave frequencies.

\section{On-Chip Wideband Sliding Correlator Channel Sounder Baseband Design }

\subsection{Principles of sliding correlator channel sounder}

\begin{figure}
\centering
\includegraphics[width=0.95\linewidth]{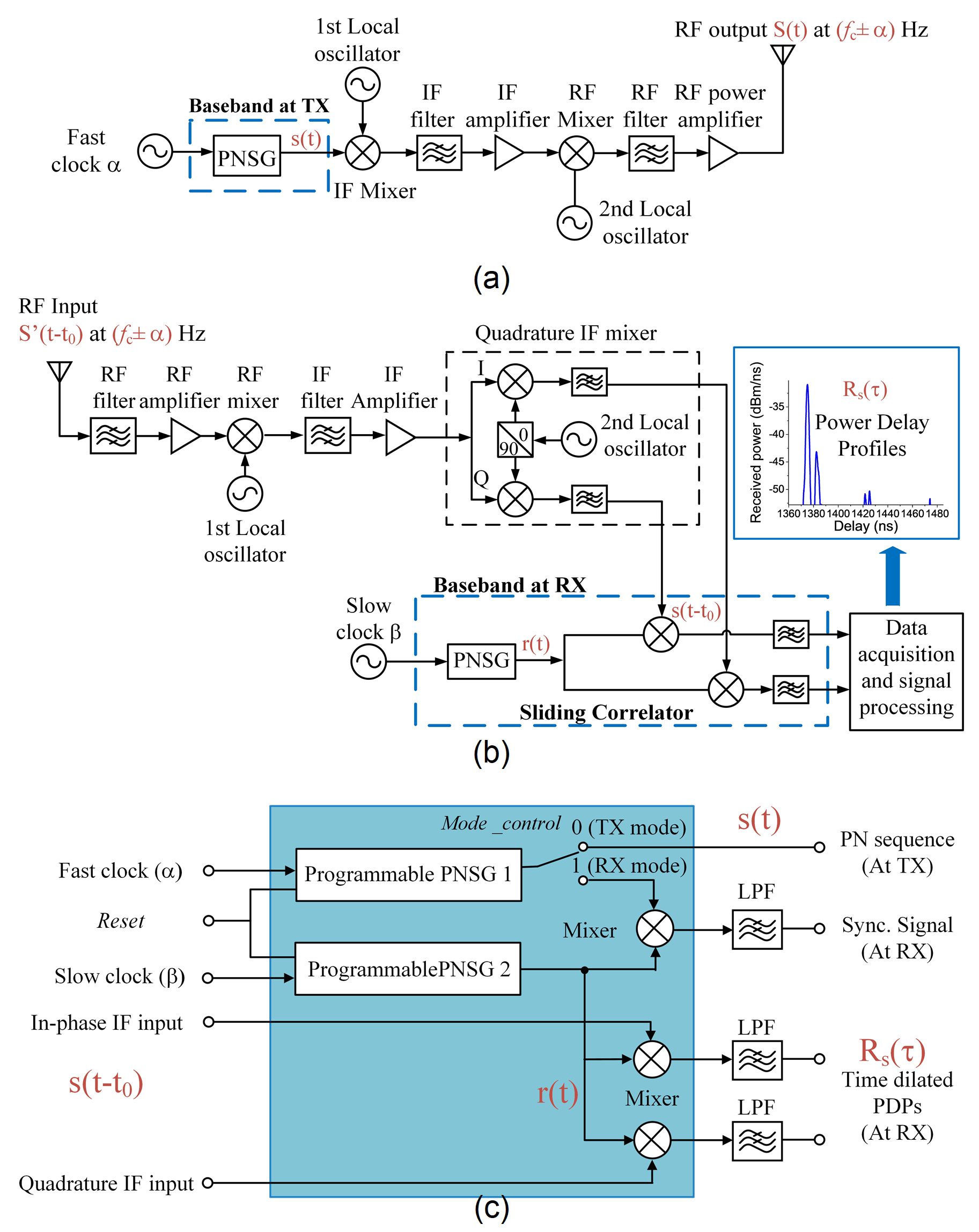}
\caption{(a) A block diagram of a typical transmitter with high chip rate pseudo-random sequence generator. (b) A block diagram of a typical receiver with sliding correlator for channel sounding applications. (c) Our proposed on-chip wideband sliding correlator-based channel sounder.  }
\label{fig:Slidingcorrelator}
\end{figure}

Fig. \ref{fig:Slidingcorrelator}(a) and (b) shows a simplified block diagram of the transmitter (TX) and receiver (RX) of a typical sliding correlator channel sounder. At the TX, the baseband pseudo-random sequence generator (PNSG) generates a PN sequence, $s(t)$, operating at the fast clock rate of $\alpha$. The maximum chip rate of the PNSG is an important parameter since it relates to the multipath delay resolution of $1/\alpha$.  Referring to Fig.\ref{fig:Slidingcorrelator}(a), the PN sequence, $s(t)$, is modulated at the intermediate frequency (IF) and then upconverted to the RF carrier frequency, $f_c$, and transmitted into the wireless channel. As the wideband transmit signal propagates through the wireless channel, it experiences reflection, scattering and diffraction effects from the environment. These ``multipath" components arrive at the RX with different propagation delays. 

At the RX, the propagation delays of the sounding signal are detected by sliding correlating using the same PNSG, but operating with a slightly lower chip rate $\beta$, as shown in Fig. \ref{fig:Slidingcorrelator}(b). The PNSG output is referred as $r(t)$. This sliding correlation can be expressed by \cite{rappaport1996wireless}

\begin{equation}\label{eq:sliding_correlation}
R_s({\tau})=\int_{0}^{T}r(t)s(t-t_0-{\tau})dt
\end{equation}

\noindent The above expression can be implemented by a pair of mixers followed by  two low pass filters. The correlated output, $R_s({\tau})$, is the power delay profile (PDP), representing the time-domain response of the measured wireless channel.  The sliding correlation dilates the time scale of the channel response measured at the receiver by a sliding factor $\gamma$, defined as \cite{rappaport1996wireless}

\begin{equation}\label{eq:sliding factor}
\gamma=\frac{\alpha}{\alpha-\beta}
\end{equation}

\noindent When examining the received signal in the frequency domain, the bandwidth is compressed, thus easing the hardware design for analog-to-digital converters. 


To extract the absolute path delay of the multipath components, it is critical to provide an output signal for synchronization at the RX baseband, i.e., a timing reference that provides a zero multipath delay. Circuits that provide sliding correlation without cable connections between the TX and RX, have difficulty in synchronizing the RX code to the TX code in the presence of numerous multipath components. One synchronization method is to use the strongest peak in the PDP as the timing reference. This method often results in excess delay relative to the strongest PDP rather than providing the desired absolute timing \cite{rappaport1996wireless,Samimi2016MTT}.


\begin{figure} [t]
\centering
\includegraphics[width = 0.80\linewidth]{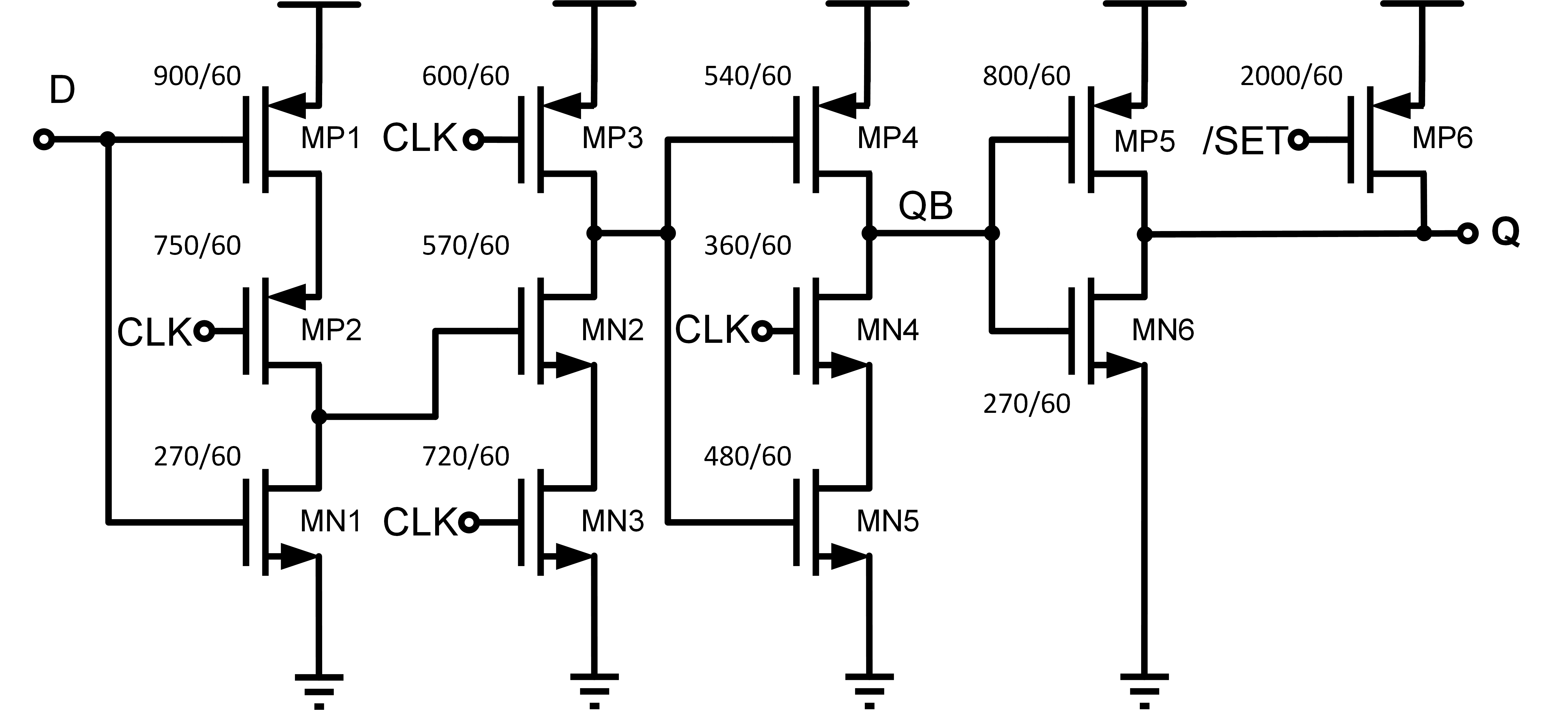}
\caption{Schematic of the positive-edge triggered TSPC DFF with low-level effective SET (/SET) based on TSPC DFF.}
\label{fig_TSPC}
\end{figure}

\subsection{Proposed on-chip baseband sliding correlator channel sounder}
Fig. \ref{fig:Slidingcorrelator}(c) shows the block diagram of the proposed on-chip spread spectrum sliding correlator channel sounder. \textit{Mode\_control} is the control bit for selecting either TX mode or RX mode. When \textit{Mode\_control} is set to 0, PNSG 1 provides the spread spectrum baseband sounding signal, $s(t)$, with chip rate $\alpha$. When \textit{Mode\_control} is set to 1, PNSG 2 provides the correlation signal, $r(t)$, with chip rate $\beta$. \textit{Mode\_control} set to 1 also provides three output signals to the RX processing system, namely \textit{Sync}, and the real and imaginary components of $R_s(\tau)$. The \textit{Sync} is a low pass filtered version of the original $s(t)$ multiplied by $r(t)$ and will be used as the absolute timing reference for the measurement. $R_s(\tau)$ is the correlation of $r(t)$ with the demodulated signal $s(t-t_0)$ where $t_0$ represents  a delay path through the wireless channel. The resulting PDPs can be time aligned and averaged, resulting in the absolute multipath delay for different multipath components. To further improve the accuracy in the absolute delay measurements, high stability clocks, such as Cesium or Rubidium standards, can be used at the baseband in the TX and RX. After calibrating the clock frequency and synchronizing the TX and RX PN codes at start of each measurement, the baseband sliding correlator channel sounder IC can provide accurate PDPs with an absolute timing reference.

\begin{figure*}[t!]
\centering
\includegraphics[width = 0.9\textwidth]{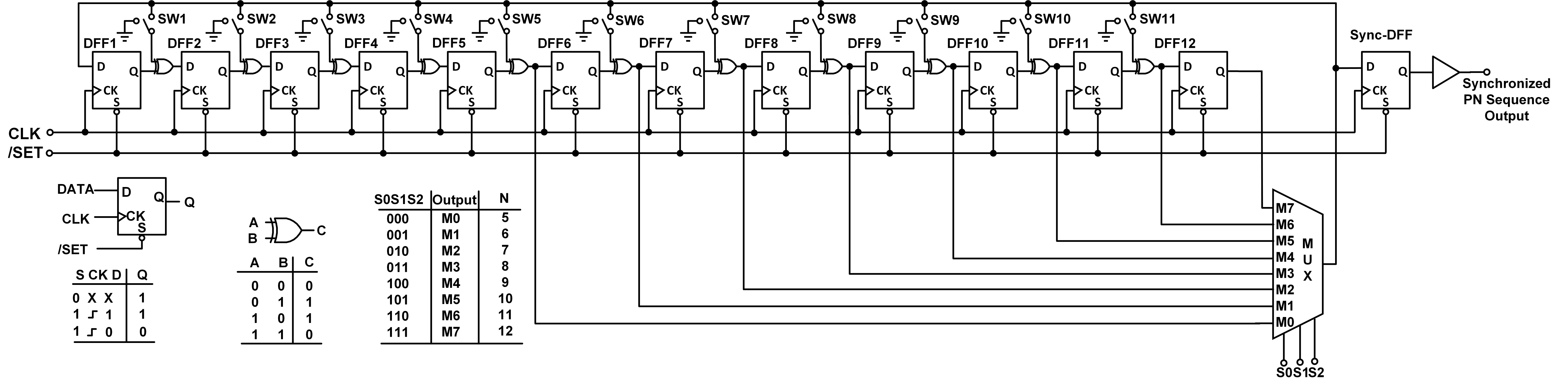}
\caption{Schematic of the proposed programmable PN sequence generator.}
\label{fig_PNBS}
\end{figure*}

\subsection{Digital D-flip-flop}
Edge triggered D flip-flop (DFF) is an important building block in a PN sequence generator. In high-speed and low-power digital VLSI designs, DFFs are often implemented using dynamic logic circuits, such as clocked CMOS (C$^2$MOS), Nora dynamic CMOS, and True single-phase-clock (TSPC). TSPC DFF is an efficient and simple technique for modern communication circuit design, because it uses a single-phase clock. The TSPC DFF of Yuan and Svensson \cite{Yuan1989:1} is constructed by a P-C$^2$MOS stage, an N-precharge stage, and an N-C$^2$MOS stage. Fig. \ref{fig_TSPC} shows the schematic of the TSPC DFF with low-level effective SET, used in this work.

\subsection{PN sequence generator}
A PN sequence can be generated using a maximal linear feedback shift register, and the generated sequence is determined by the number of stages in the shift register and a specific combinations of feedback taps \cite{pirkl2008optimal}. Two types of shift register generators are commonly used: simple shift register generator (SSRG) and modular shift register generator (MSRG). In MSRG, the feedback state from the last stage is added to modulo-2 adders between each stage, and the total delay is less than that of a single-tap sequence generator. Thus, when more than one feedback connections are used, MSRG is preferable than SSRG in terms of speed \cite{Dixon94}.  

The proposed PN sequence generator based on MSRG configuration is shown in Fig. \ref{fig_PNBS}. A 3-bit digital words $S\langle2:0\rangle$  selects the input of the multiplexer to the output, thus determining the number of stages, while $SW\langle12:1\rangle$ controls the feedback taps. Digital words $SW\langle12:1\rangle$ should be carefully selected to generate a PN code.

To avoid the ``all-zero" condition, the initial state of each DFF is set to one by a low-level effective SET signal (/SET) after power up. The thermal and shot noise can cause the time of the rising edge (or the falling edge) of the binary sequences to vary, adding jitter to the signal, which can accumulate through the propagation of the signal \cite{sahafi2013ultra}. In order to minimize the jitter, the generated PN sequence from the last stage is followed by a synchronization DFF (Sync-DFF in Fig. \ref{fig_PNBS}). The Sync-DFF has the same clock signal as the other DFFs in the shift register, thus the rising edge of the signal is synchronized with the rising edge of the main clock signal CLK. This way, the jitter will be minimized because only one DFF contributes to the jitter.

\subsection{Gilbert cell double-balanced mixer}

\begin{figure}
\centering
\includegraphics[width = 1.5in]{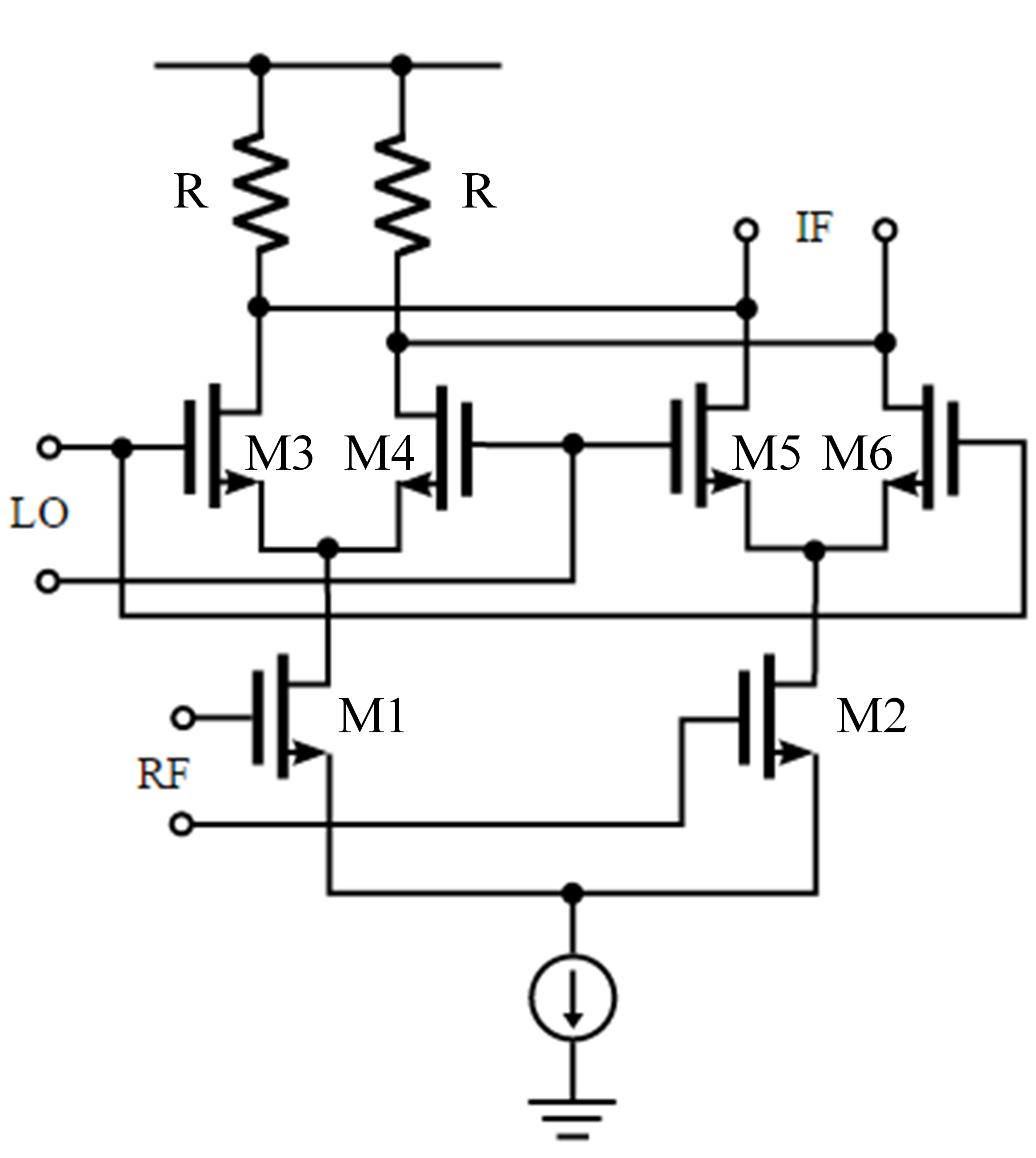}
\caption{Schematic of the Gilbert cell double-balanced mixer.}
\label{fig_mixer}
\end{figure}
In order to achieve better signal-to-noise ratio in the sliding correlator channel sounder system, the receiver front-end architecture should provide differential signaling including the mixers. In this design, the mixing in the sliding correlator IC was realized using a Gilbert cell double-balanced mixer shown in Fig. \ref {fig_mixer}. M1 and M2 convert RF input voltage to current. The other four transistors are driven by the LO signal to switch the polarity of the RF current. To improve the mixer performance, sufficient drive from the LO and appropriate biasing must be supplied in order to make the four LO transistors behave as close as possible to ideal switches.

\section{Measurement Results}

\begin{figure}
\centering
\includegraphics[width = 2.0in]{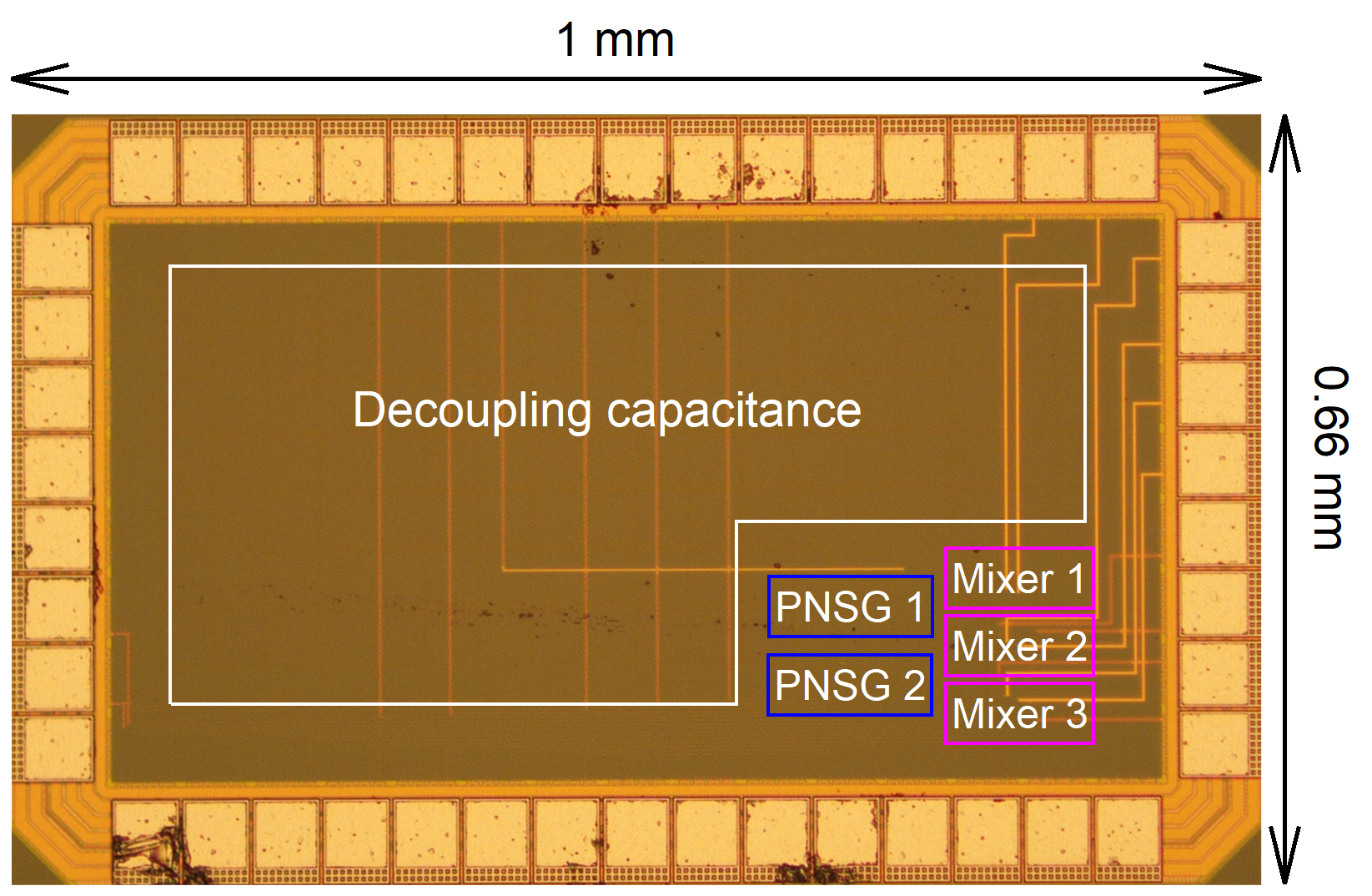}
\caption{Photograph of the sliding correlator channel sounder baseband IC. }
\label{fig_OpticalImage}
\end{figure}

The programmable spread spectrum channel sounder baseband IC was fabricated in a standard 65 nm CMOS process and powered using a 1.1 V power supply. Fig. \ref{fig_OpticalImage} shows the chip photograph. The chip occupies an area of 0.66 mm $\times$ 1 mm. 

In order to test the chip, a 4-layer PCB using a FR4 substrate was fabricated and the chip was attached to the board and wire bonded for the signal inputs and outputs. Passive low-pass filters with cut-off bandwidths of 100 kHz, were also soldered to the appropriate chip outputs.

\begin{figure}[t!]
\centering
\includegraphics[width=0.85\linewidth]{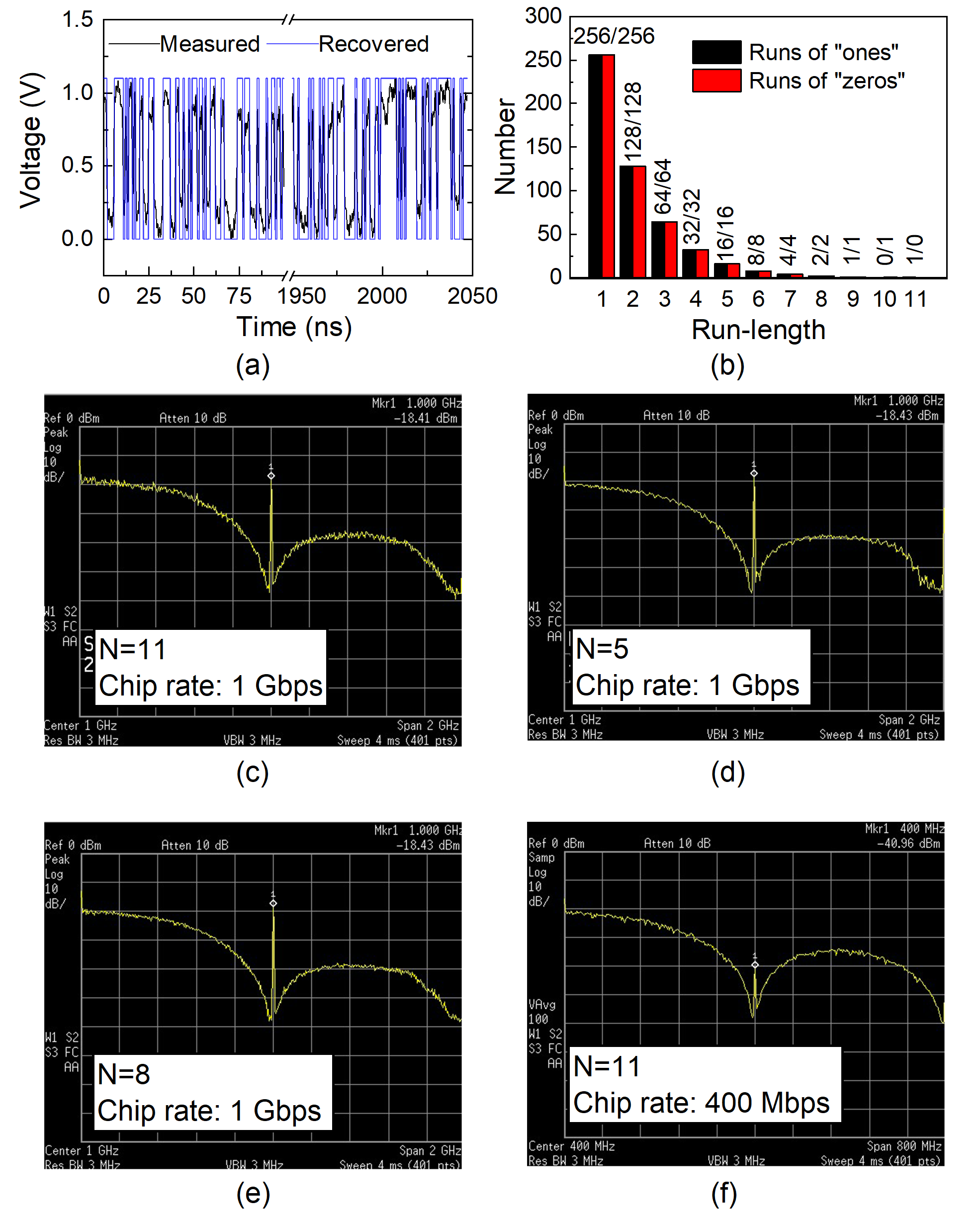}
\caption{Validation of generated m-sequence: (a) Measured and recovered of a 11-stage PNSG at 1 Gbps.(b) Distribution of runs for an 11-stage configuration at 1 GHz (c) Measured power spectrum of the 11-stage PN sequence at 1 Gbps showed in (a). (d) Measured power spectrum of a 5-stage PN sequence at 1 Gbps. (e) Measured power spectrum of an 8-stage  PN sequence at 1 Gbps. (f) Measured power spectrum of an 11-stage PN sequence at 400 Mbps. }
\label{fig_validation}
\end{figure}

Fig. \ref{fig_validation}(a) shows the time-domain measurement result from PNSG 1 (\textit{Mode\_control} = 0, TX mode). The PN sequence is generated by an 11-stage shift register with a 1 GHz clock. Controls  $S\langle{2:0}\rangle$ = ``110", and $SW\langle{12:1}\rangle$ = ``00010010010", result in feedback taps of [11,8,5,2]. In order to validate that the measured sequence was a maximal sequence (m-sequence), we first recover the measured analog signal  (black) to a binary sequence (blue). Then we count the number of various run-lengths for both ``ones" and ``zeros" groups (One run is defined as a series of ``ones" or ``zeros" grouped consecutively), shown in Fig. \ref{fig_validation}(b). The measured PN sequence has only one run containing eleven ``ones" and only one run containing ten ``zeros" and equal number of ``ones" and ``zeros" for the other run-lengths, confirming it is an m-sequence code \cite{Dixon94}. The measured spectrum for this PN sequence operating with a clock frequency of 1 GHz is shown in Fig. \ref{fig_validation}(c). This baseband chip has the capability to program the number of shift register stages in the m-sequence PN code. Fig. \ref{fig_validation}(d) and (e) shows the measured spectrum for a 5-stage and 8-stage PN code clocked at 1 GHz.  Fig. \ref{fig_validation}(f) shows the spectrum for the 11-stage PN code operating with a 400 MHz clock. This measurement confirms that our baseband IC is flexible and can achieve a maximum clock rate of 1 GHz while also providing programmability in output sequence.

\begin{figure}[t!]
\centering
\includegraphics[width = 0.6\linewidth]{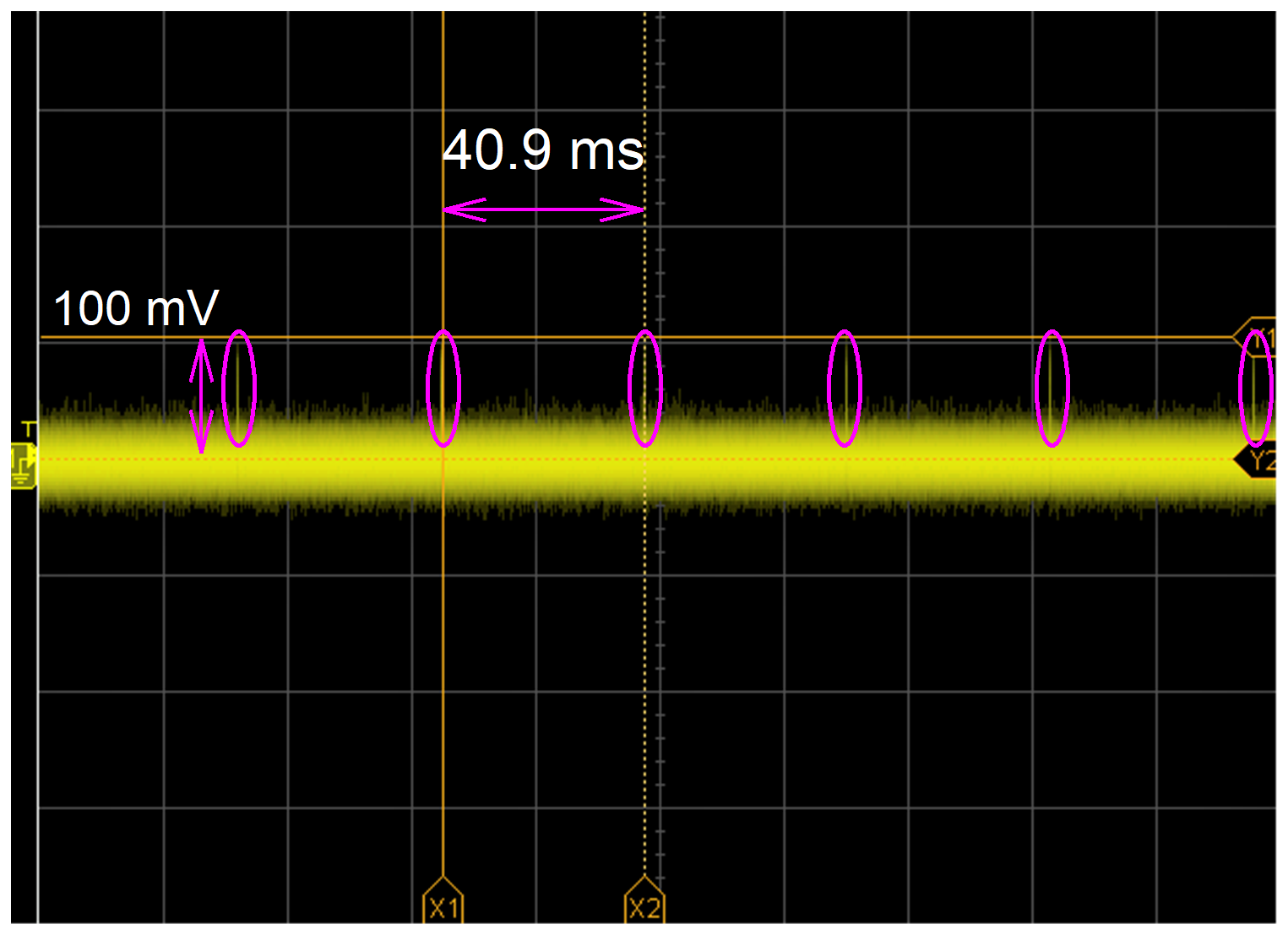}
\caption{Measurement of the synchronization signal which can be used as a timing reference to extract absolute multipath delay of the PDPs. }
\label{fig_Sync}
\end{figure}

Fig. \ref{fig_Sync} shows the measurement results of the synchronization signal generated when \textit{Mode\_control} = 1 (RX mode). Two identical 11-stage PN sequences, clocked at 1 GHz and 999.95 MHz respectively, are correlated together to generate the required synchronization signal. Since the length of PN sequence is 2047 chips, the synchronization signal repeats every 40.94 ms. The amplitude of the measured  synchronization signal is about 100 mV, which could be further increased if the passive low-pass filters are replaced with active filters.  

Table \ref{table1} is a summary of key specifications of the channel sounder baseband chip.

\begin{table} [t!]
\renewcommand{\arraystretch}{1.2}
\caption{A summary of specifications of our on-chip baseband design for wideband sliding correlator channel sounder.}
\label{table1}
\centering
\begin{tabular}{|>{\centering}p{3.5cm}|>{\centering}p{3.5cm}|}
\hline

  {\textbf{Item}}&{\textbf{Specification}}\tabularnewline
\hline

 \textbf{Technology}& 65 nm CMOS \tabularnewline
\hline
\textbf{Maximum chip rate}& 1 Gbps \tabularnewline
\hline
\textbf{Multipath delay resolution} & 1 ns \tabularnewline
\hline
\textbf{Null-to-null RF bandwidth} & 2 GHz \tabularnewline
\hline
\textbf{PN code length} & $2^N-1$ (N is programmable from 5 to 12) \tabularnewline
\hline
\textbf{Area} & 0.66 mm $\times$ 1 mm \tabularnewline
\hline
\textbf{Synchronization} & Supported \tabularnewline
\hline

\end{tabular}
\label{table1}
\end{table}

\section{Conclusion}

In this paper, we presented an on-chip ultra-wideband spread spectrum sliding correlator-based channel sounder with high spatiotemporal resolution. This work is the first known report to achieve a 1 Gbps baseband operation using low-cost CMOS technology. The versatility of this channel sounder provides researchers with the capability to choose the most suitable system parameters when conducting propagation measurements with high accuracy and efficiency. The miniaturized baseband chip can be used for accurate spatiotemporal channel propagation measurements, to identify absolute multipath delays, in system utilizing massive MIMO technology at millimeter-wave carrier frequencies. 

\section*{Acknowledgement}

The authors acknowledge United Microelectronics Corporation (UMC) for manufacturing the chip. This research is supported by the NYU WIRELESS Industrial Affiliates Program.

\bibliographystyle{IEEEtran}


\end{document}